\documentclass[letter,fleqn,usenatbib]{mnras}

\usepackage[T1]{fontenc}
\usepackage{ae,aecompl}

\usepackage{graphicx}	
\usepackage{amsmath}	
\usepackage{amssymb}	

\usepackage{times}
\usepackage{epstopdf}
\usepackage{graphicx}
\usepackage{txfonts}
\usepackage{natbib}
\usepackage{array}
\title[ALMA imaging of SDP.81 -- II.]{ALMA imaging of SDP.81 -- II.  A pixelated reconstruction of the CO emission lines}
\author[M. Rybak et al.]{M. Rybak,$^{1}$\thanks{E-mail: mrybak@mpa-garching.mpg.de} S. Vegetti,$^{1}$ J. P. McKean,$^{2,3}$ P. Andreani$^{4}$ and S. D. M. White$^{1}$\\
$^{1}$Max Planck Institute for Astrophysics, Karl-Schwarzschild-Strasse 1, 85740 Garching, Germany\\
$^{2}$Netherlands Institute for Radio Astronomy (ASTRON), P.O. Box 2, 7990 AA Dwingeloo, The Netherlands\\
$^{3}$Kapteyn Astronomical Institute, University of Groningen, P.O. Box 800, 9700 AV Groningen, The Netherlands\\
$^{4}$European Southern Observatory, Karl-Schwarzschild-Strasse 2, 85748 Garching, Germany}

\begin{document}

\date{Accepted 2015 July 6.  Received 2015 June 27; in original form 2015 June 2}

\pagerange{\pageref{firstpage}--\pageref{lastpage}} \pubyear{2015}

\maketitle

\label{firstpage}
\begin{abstract}
We present a sub-100~pc-scale analysis of the CO molecular gas emission and kinematics of the gravitational lens system SDP.81 at redshift 3.042 using Atacama Large Millimetre/submillimetre Array (ALMA) science verification data and a visibility-plane lens reconstruction technique. We find clear evidence for an excitation dependent structure in the unlensed molecular gas distribution, with emission in CO (5-4) being significantly more diffuse and structured than in CO (8-7). The intrinsic line luminosity ratio is $r_{\rm 8-7 / 5-4} =$~0.30\,$\pm$\,0.04, which is consistent with other low-excitation starbursts at $z \sim$~3. An analysis of the velocity-fields shows evidence for a star-forming disk with multiple velocity components that is consistent with a merger/post-coalescence merger scenario, and a dynamical mass of $M$($<$\,1.56~kpc) = 1.6\,$\pm$\,0.6\,$\times$\,10$^{10}$~M$_{\odot}$. Source reconstructions from ALMA and the {\it Hubble Space Telescope} show that the stellar component is offset from the molecular gas and dust components. Together with Karl G. Jansky Very Large Array CO (1-0) data, they provide corroborative evidence for a complex $\sim$2~kpc-scale starburst that is embedded within a larger $\sim$15 kpc structure.
\end{abstract}


\section{Introduction}
Recent large-area surveys have revealed a population of dusty sub-millimetre (sub-mm) galaxies at high redshift \citep[e.g.][]{greve04,weiss09} that are believed to be the progenitors of  present-day massive early-type galaxies \citep{Swinbank06}. These galaxies are characterized by their extreme far-infrared (FIR) luminosities, considerable stellar masses and large star-formation rates (e.g. \citealt{coppin08}). On the basis of morphological and kinematic studies at ultraviolet (UV) and optical wavelengths, it has been suggested that most of these galaxies are undergoing a major merger event \citep[e.g.][]{Swinbank06,Engel2010}. In this scenario, their extreme luminosities are the result of merger-induced star formation. A competing mechanism, in which star formation is predominantly driven by in-falling gas-rich material has also been proposed (e.g. \citealt*{dekel09a,carilli10}). 

By using high angular-resolution observations of the molecular gas component of these galaxies, it is possible to constrain their kinematic and morphological properties, and establish the origin of their intense star-formation in a way that is neither affected by dust extinction nor biased by outflows \citep{Tacconi2008,Engel2010,Karim2013}. While most of these studies have so far been hampered by the limited spatial resolution of the observations, the advent of the Atacama Large Millimetre/submillimetre Array (ALMA) has made it possible to make sensitive studies of sub-mm galaxies on arcsec-scales \citep{Karim2013}. The recent addition of the long baseline capacity of ALMA has further improved the imaging quality (30 to 300 mas resolution; e.g. \citealt{alma15a}). At the same time, strong gravitational lensing has allowed the study of high-redshift sub-mm galaxies on 50-100~pc-scales, which is comparable to what is achieved at lower redshift \citep{swinbank10}.

In this Letter, we take advantage of the enhanced spatial resolution provided by the long baseline capability of ALMA, in combination with the magnifying power of strong gravitational lensing to study the molecular gas distribution and kinematics of the starburst galaxy SDP.81 (H-ATLAS J090311.6+003906; $z =$ 3.042\,$\pm$\,0.001; \citealt{Negrello2010,Negrello14,Frayer2011,Bussmann2013,Dye14}). In a previous paper (\citealt{Rybak15a}; referred to as Paper I), we presented an analysis of the gravitational lens mass model and a study of the reconstructed heated dust continuum emission at a resolution better than 50~pc. Here, we focus on a detailed analysis of the CO emission line data that were also taken during the science verification of the ALMA long baseline array. These data are coupled with high angular resolution imaging from the {\it Hubble Space Telescope} ({\it HST}) and Karl G. Jansky Very Large Array (VLA) to investigate the conditions that produced the intense burst of star-formation within SDP.81.

\section{Observations \& lens modelling}

\subsection{ALMA CO emission line imaging}
\label{ALMA}

The ALMA spectral line observations of SDP.81 and the image processing are discussed in detail by the \citet{alma15b}. To summarize, the CO (5-4), CO (8-7), CO (10-9) and H$_2$O (2$_{02}$-1$_{11}$) transitions were observed for between 4.4 and 5.9~h in Bands 4, 6 and 7. The array consisted of between 22 and 36 antennas with baselines ranging from 15 m to 15 km. During our analysis, we found that the CO (10-9) and H$_2$O emission was too weak to provide reliable velocity and velocity dispersion maps, so we do not consider them further here, and instead we focus on the CO (5-4) and CO (8-7) transitions.

The gravitational lens mass model for SDP.81 was derived in Paper~I by applying a Bayesian pixelated modelling technique (Rybak et al., in prep.) to the visibility data of the Band 6 and 7 continuum emission. This lens model is used for the analysis of the CO emission line data here, that is, the lens parameters are kept fixed at the most probable values derived in Paper~I, while the emission line structure and its regularization level are left as free parameters. In order to model the two-dimensional spectral maps of the CO lines we first subtracted the continuum emission from the {\it uv}-dataset using the line-free channels. As the emission lines are quite broad, we averaged the data by 2 channels resulting in a velocity resolution lines of 41~km\,s$^{-1}$ for both lines. This was found to be a reasonable compromise between velocity space sampling (to study the structure of the lines) and signal-to-noise ratio (SNR) for a robust source reconstruction. Also, to limit the number of visibilities used in our analysis and to ensure that we compare the various transitions on the same angular-scale, a 1~M$\lambda$ cut in the {\it uv}-distance was imposed. For a Briggs weighting scheme with Robust = 0, this corresponds to a synthesised beam with a full width at half-maximum (FWHM) of 180\,$\times$\,131~mas at a position angle of 77.7~deg.

\subsection{Archival {\it HST}-WFC3 infrared imaging}
\label{HST}

\begin{figure}
\begin{center}
{\includegraphics[height=6.0cm, clip=true]
{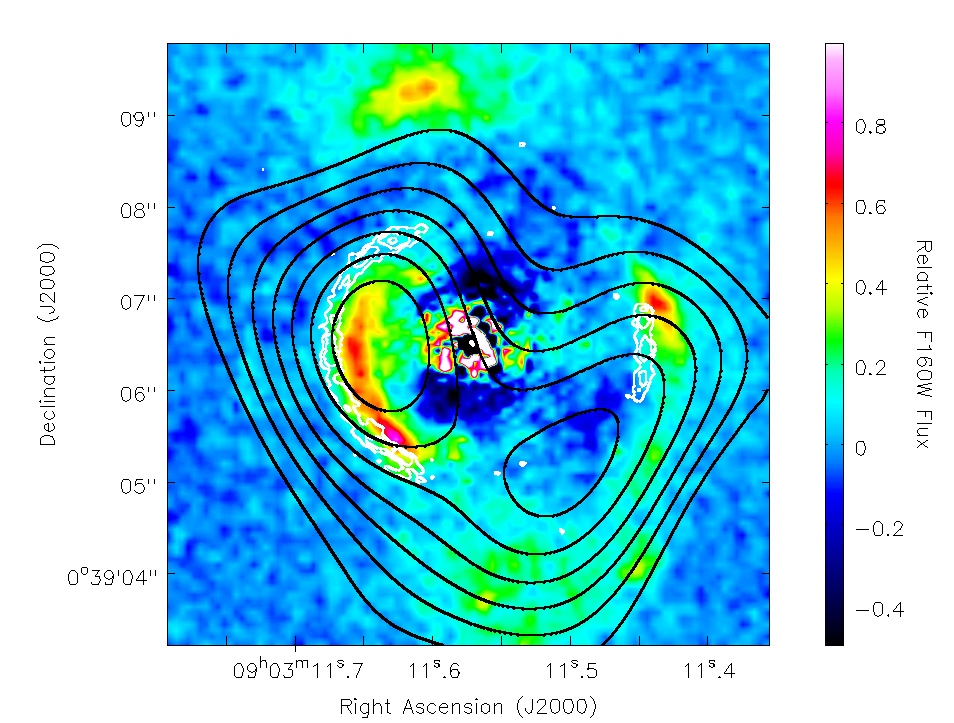}}
\caption{The image-plane {\it HST} (WFC3 F160W) rest-frame UV/optical emission from SDP.81 (lensing galaxy subtracted) with the ALMA dust-continuum (white contours) and VLA CO (1-0) emission (black contours; at 0.3, 0.4, 0.5, 0.6, 0.7, 0.8, 0.9 relative to the peak emission). For the CO (5-4) and (8-7) molecular gas distributions, see \citet{alma15b}.}
\label{fig:image}
\end{center}
\end{figure}

SDP.81 was observed with the Wide Field Camera 3 aboard the {\it HST} on 2011 April 4 (GO: 12194; PI: Negrello). The data were taken through the F110W and F160W filters, with exposure times of 712 and 4418~s, respectively (see \citealt{Negrello14} for further details). The {\it HST} imaging detected a single dominant lensing galaxy, two gravitational arcs from the background source and extended low-surface brightness structures around the system. 

The data were reprocessed using the standard {\sc astrodrizzle} pipeline (using a pixel size of 65~mas) and the subtraction of the lensing galaxy emission was done using {\sc galfit} \citep{peng10}. The image-plane structure of the {\it HST} and ALMA continuum emission (the latter from Paper I) are shown in Fig.~\ref{fig:image}. Although the arcs are offset, due to their different positions in the source-plane (see below), the separation between the main and counter-arc is the same in both the infrared and mm components, which demonstrates that they are at the same redshift. We model the infrared emission using the pixelated gravitational lensing code of \citet{Vegetti09}. Consistent with previous studies (see \citealt{Dye14,dye15,tamura15,wong15}), we find that the infrared emission has a magnification of $\mu_{\rm tot}=$~11.3\,$\pm$\,0.1. We discuss the source-plane properties of the rest-frame UV/optical emission, with respect to the dust and gas components, in Section~\ref{conc}.

\subsection{Archival VLA imaging}
\label{VLA}

Spectral line imaging of the CO~(1-0) emission from SDP.81 was carried out using the VLA in D-configuration on 2010 July 18 (PI:~Ivison; 3.3\,$\times$\,2.3~arcsec  beam-size at position angle $-$10.1 degr). The integration time was 1.5~h and 16 VLA antennas that had been upgraded with the new receiver systems were used. Further details about these observations are describe by \citet{valtchanov11}. We obtained these data from the VLA archive and reprocessed them in the standard manner using the Common Astronomy Software Applications ({\sc casa}) package. In Fig.~\ref{fig:image}, we show the image-plane emission of the CO~(1-0) transition, with respect to the infrared and mm-continuum emission (see above and Paper I). It is clear that the CO (1-0) emission is co-located with the infrared-arcs and low-surface brightness infrared emission to the south, demonstrating that these extended infrared components are at the same redshift as SDP.81. The SNR of the VLA dataset is too low to provide a reliable source reconstruction using our visibility lens modelling technique or to determine if the infrared component to the north of SDP.81 is also associated with the lens system.

\section{Reconstructions \& Analysis}
\label{results}

\begin{figure}
\begin{center}
{\includegraphics[height=3.0cm, clip=true]
{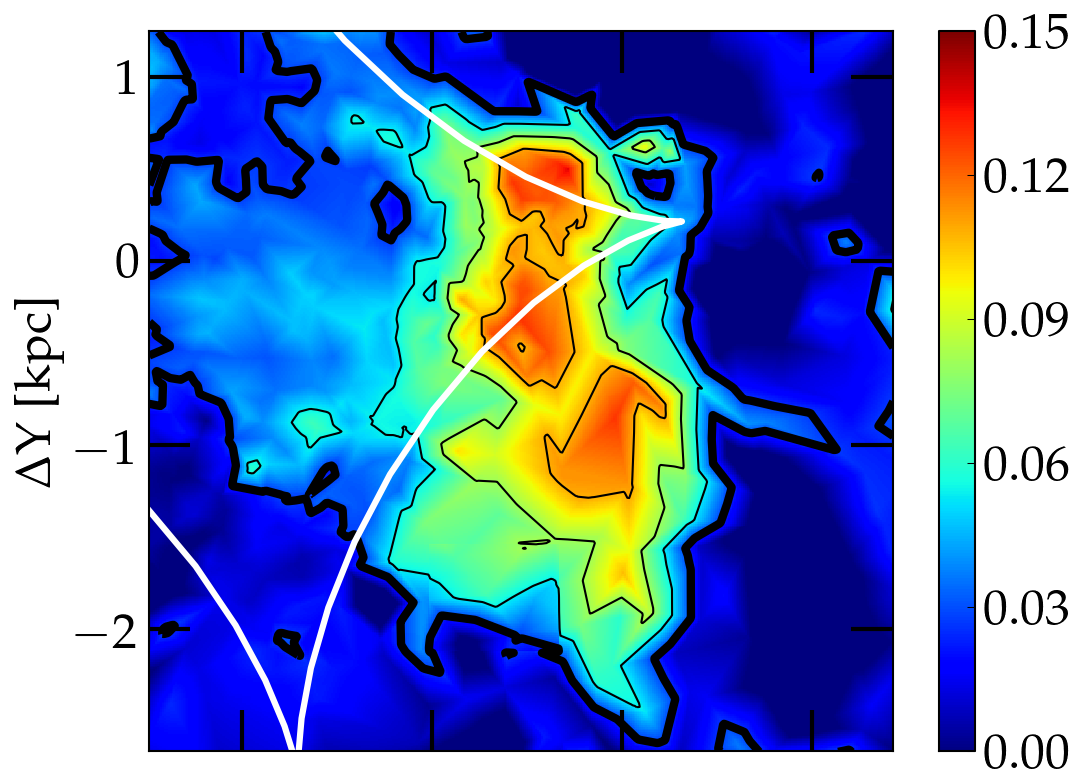}}
{\includegraphics[height=3.0cm, clip=true]
{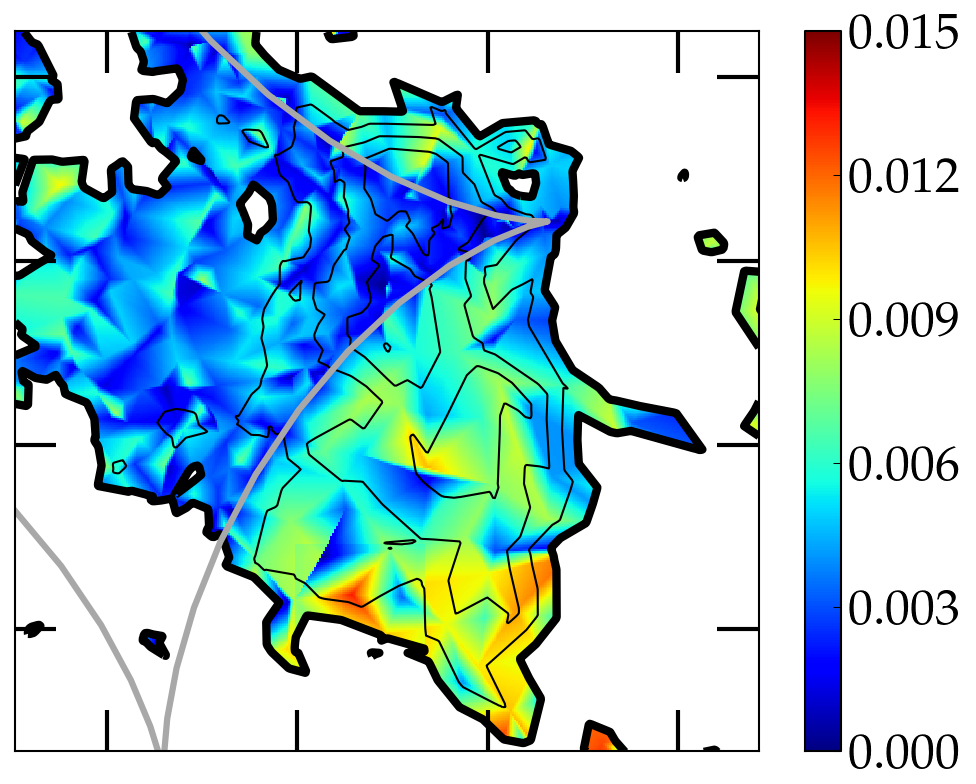}}\\
{\includegraphics[height=3.33cm, clip=true]
{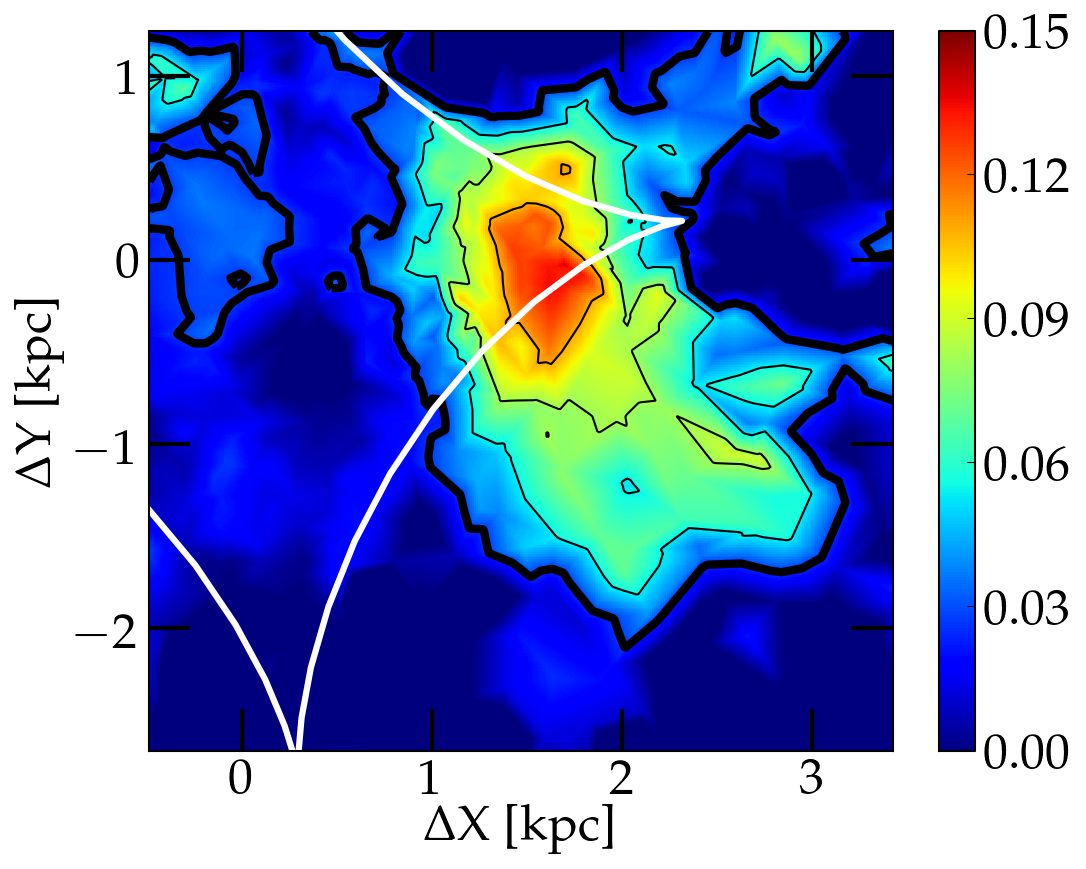}}
{\includegraphics[height=3.33cm, clip=true]
{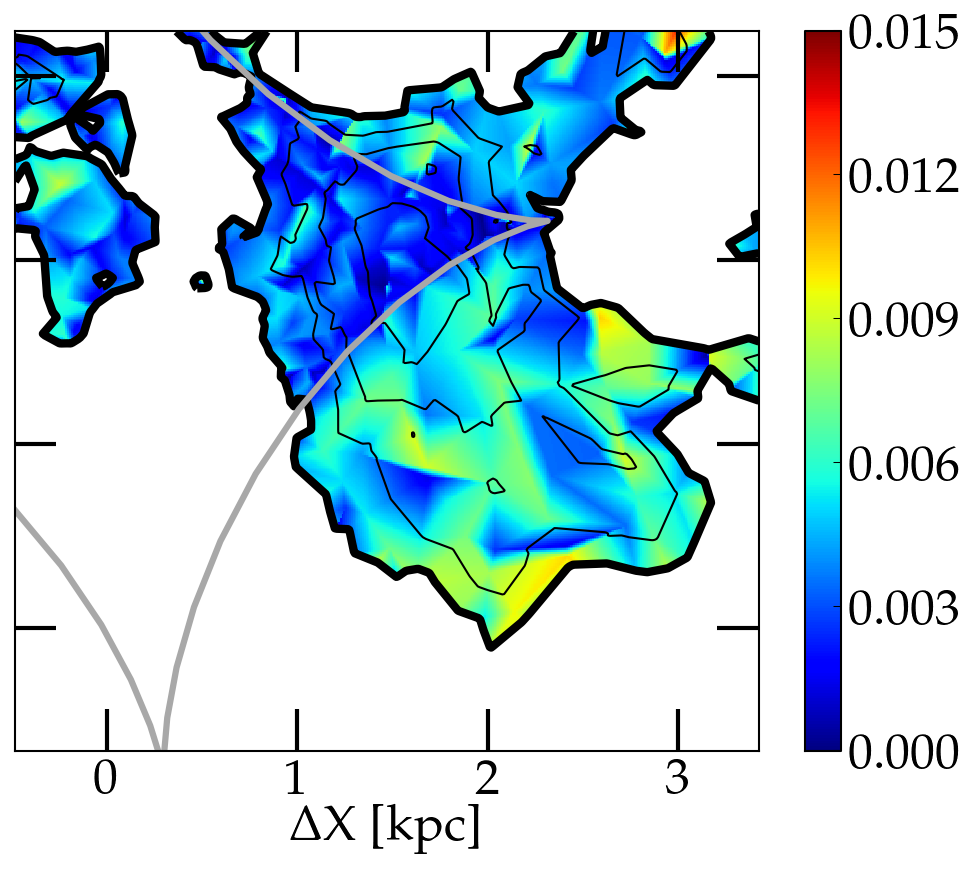}}
\caption{The zeroth-moment (left) and uncertainty (right) maps for the (upper) CO (5-4) and (lower) CO (8-7) emission lines. The colour-bar is in units of Jy~km\,s$^{-1}$~kpc$^{-2}$. The lens centre is at co-ordinate (0, 0) and the source-plane caustic is shown by the (left) white and  (right) grey lines. The contours are drawn at 0.2, 0.4, 0.6 and 0.8 relative to the peak emission.}
\label{fig:moment0}
\end{center}
\end{figure}

The source-plane velocity integrated emission of the CO (5-4) and CO (8-7) transitions from ALMA are presented in Fig.~\ref{fig:moment0}. These maps show clear excitation dependent structure in the molecular gas distribution of SDP.81. The CO (5-4) emission has both diffuse and multiple compact structures that extend over $\sim$3~kpc in the north-south direction, with an additional low surface brightness component that is $\sim$1--2~kpc in size towards the north-east of the source. The emission from the CO (8-7) transition is much more compact, with an extent of about $\sim$1.5~kpc, and there is only a single brightness clump. There is also evidence of a faint tail of emission that extends by $\sim$2~kpc to the south-west; this tail is not seen in the CO (5-4) transition, which implies there is higher excitation gas in that region, possibly due to shocks. 

As a check, we compare our reconstructions with the image-plane zeroth-moment maps of the counter-arc \citep{alma15b}; the counter-arc does not cross a critical curve (see Paper~I) and so the magnification is roughly constant across this image. We find that our source-plane reconstructions are in good agreement with the observed image-plane structure of the counter-arc. The total magnification of the CO (5-4) and CO (8-7) gas components are 17.0\,$\pm$\,0.4 and 16.9\,$\pm$\,1.1, respectively.

The total velocity integrated emission from the CO (5-4) and CO (8-7) transitions has been determined by directly integrating the zeroth-moment maps over the total extent of the reconstructed source. We find the velocity integrated line intensities are 0.54\,$\pm$\,0.06 and 0.45\,$\pm$\,0.03~Jy~km\,s$^{-1}$ for the CO (5-4) and CO (8-7), respectively. These correspond to line luminosities of $L'_{\rm CO\,(5-4)}=$~0.91\,$\pm$\,0.10\,$\times$\,10$^{10}$~K~km\,s$^{-1}$~pc$^2$ and $L'_{\rm CO\,(8-7)}=$~0.27\,$\pm$\,0.02\,$\times$\,10$^{10}$~K~km\,s$^{-1}$~pc$^2$. This gives an intrinsic line luminosity ratio of $r_{\rm 8-7 / 5-4} =$~0.30\,$\pm$\,0.04, which is consistent with other starburst galaxies at redshift 2--3 that typically show low excitation \citep[e.g.][]{Scott11}. 

Ideally, we would like to compare with the line luminosity of the CO (1-0) emission. However, simply applying our magnification of the CO (5-4) or CO (8-7) transitions to the observed CO (1-0) spectrum would be incorrect; the CO (1-0) line is offset from the dust continuum emission and so does not have the same magnification (see Fig.~\ref{fig:image}). Also, the CO (1-0) emission from starburst galaxies has been found to be extended by up to $\sim$16~kpc (FWHM; \citealt{ivison11}), and so the magnification of this gas component is likely much smaller than for the CO (5-4) or CO (8-7) emission.

\begin{figure}
\begin{center}
{\includegraphics[height=3.3cm, clip=true]
{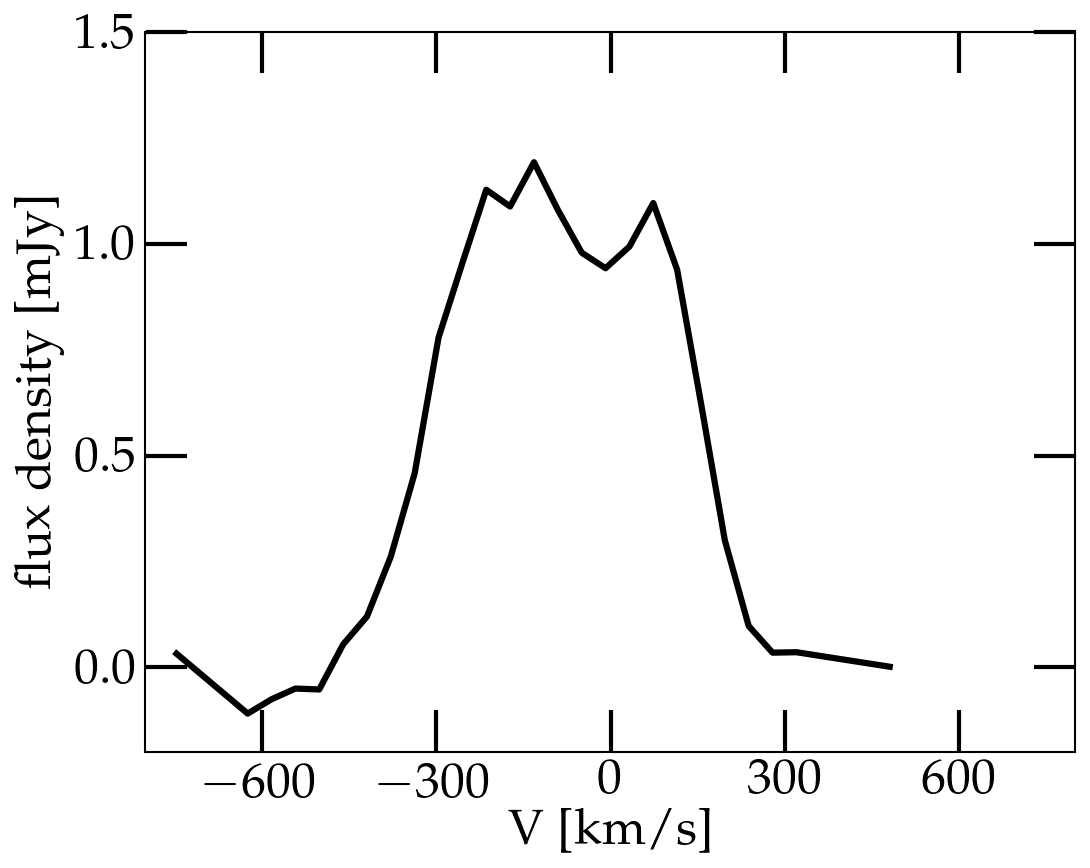}}
{\includegraphics[height=3.3cm, clip=true]
{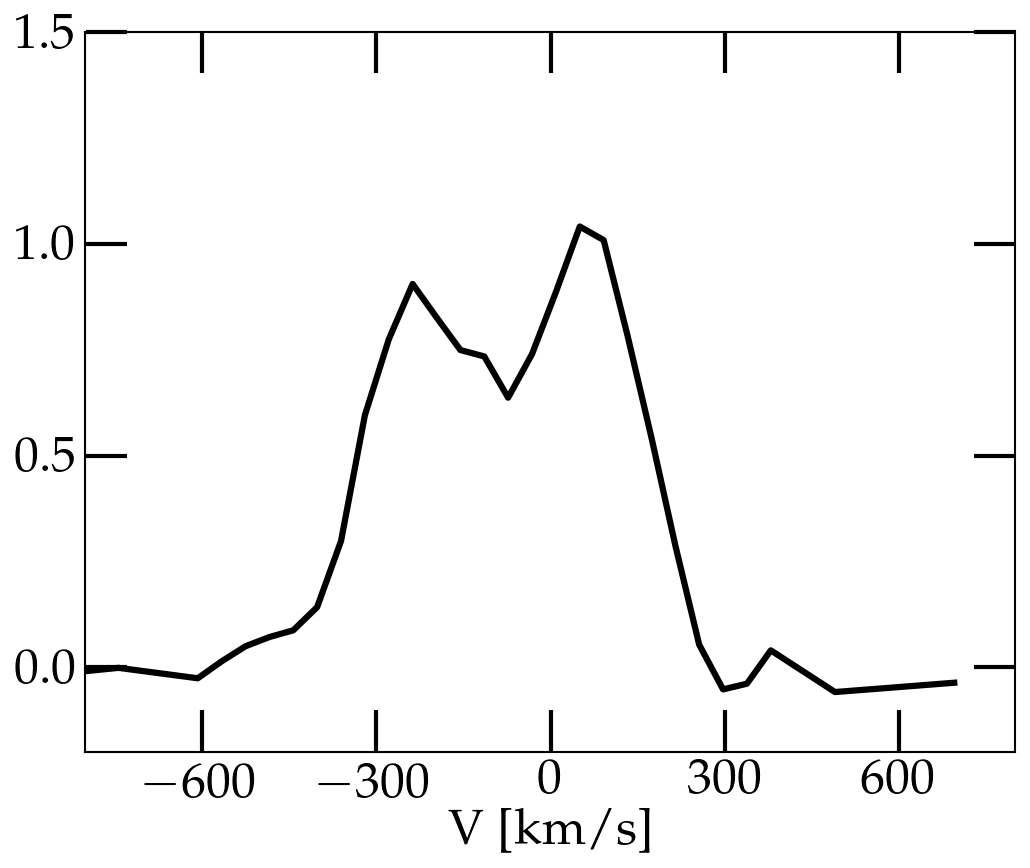}}
\caption{The intrinsic line profiles of the (left) CO (5-4) and (right) CO (8-7) transitions, smoothed with a boxcar of 120 km\,s$^{-1}$ width, and assuming a systemic velocity corresponding to a redshift of 3.042 and using the radio definition of the velocity.}
\label{fig:profiles}
\end{center}
\end{figure}

The spatially dependent structure of the CO emission lines extends over the caustic curves, which explains the complex velocity structure that has been observed in the main arc of SDP.81 (see \citealt{alma15b}); for example, the image-plane line profile of both transitions are highly asymmetric. In Fig.~\ref{fig:profiles}, we present the intrinsic line profiles of both the CO (5-4) and CO (8-7) transitions that are corrected for the differential magnification. We find that both transitions actually have more symmetric profiles, consisting of a double-horn profile, typical of rotating disks, with velocity peaks that are separated by $\sim$\,290~km\,s$^{-1}$. Again, this result is consistent with the image-plane line profiles of the counter arc. The CO (5-4) and CO (8-7) emission lines are also offset by about $-$80~km\,s$^{-1}$ from the systemic velocity, as defined by the CO (1-0) measurement \citep{Frayer2011}. This could be due to there being turbulent in/out flowing higher excitation gas, or that the CO (1-0) line also undergoes differential magnification.

\begin{figure}
\begin{center}
{\includegraphics[height=3.0cm, clip=true]
{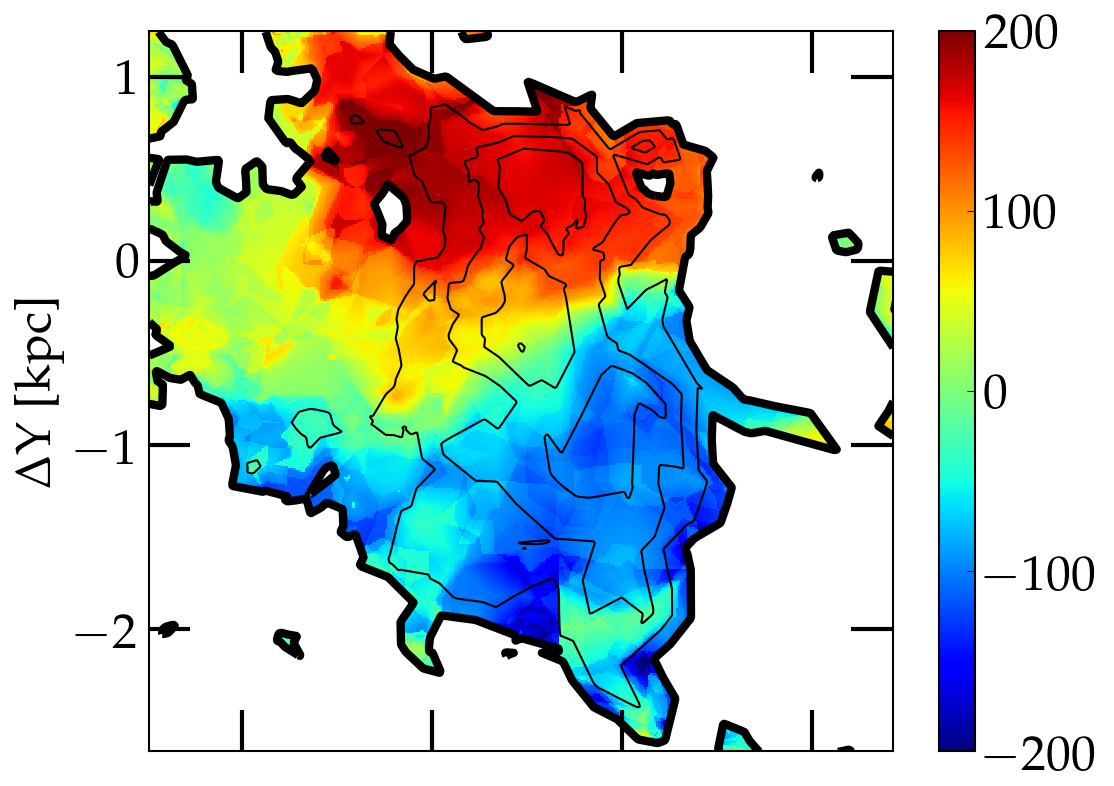}}
{\includegraphics[height=3.0cm, clip=true]
{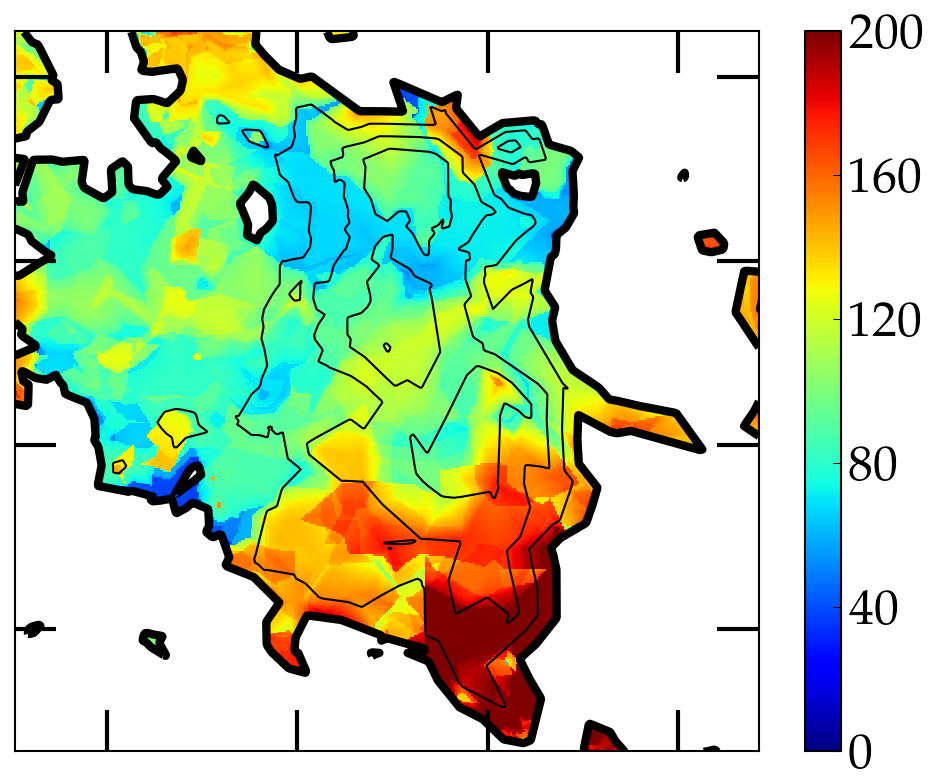}}\\
{\includegraphics[height=3.0cm, clip=true]
{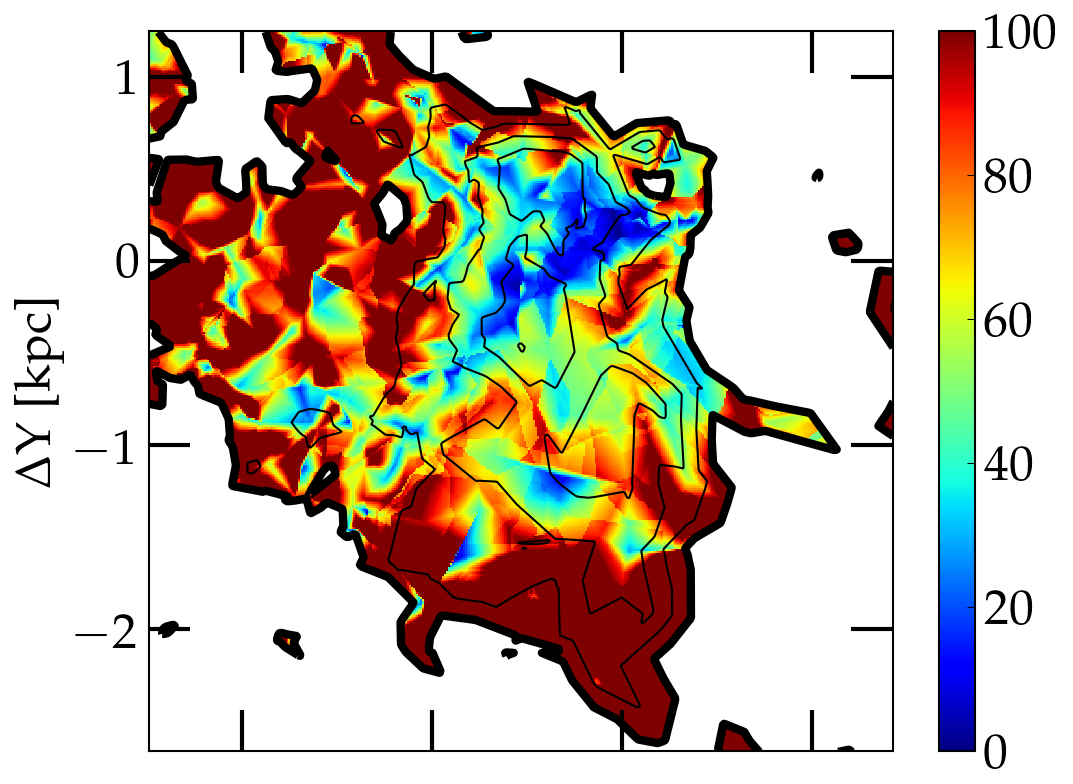}}
{\includegraphics[height=3.0cm,clip=true]
{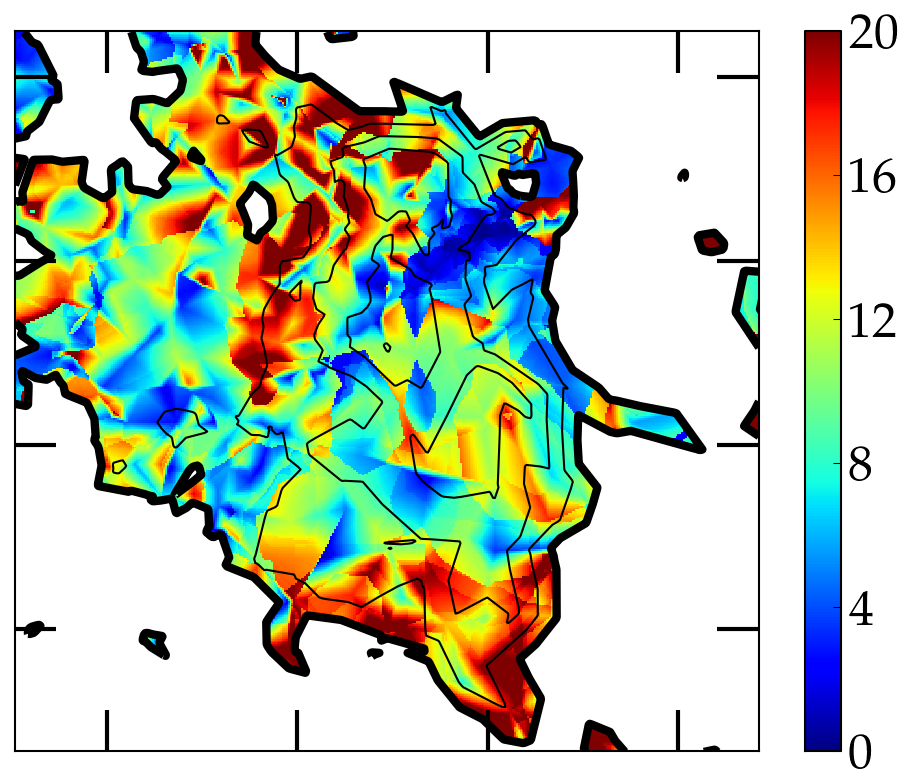}}\\
{\includegraphics[height=3.33cm, clip=true,]
{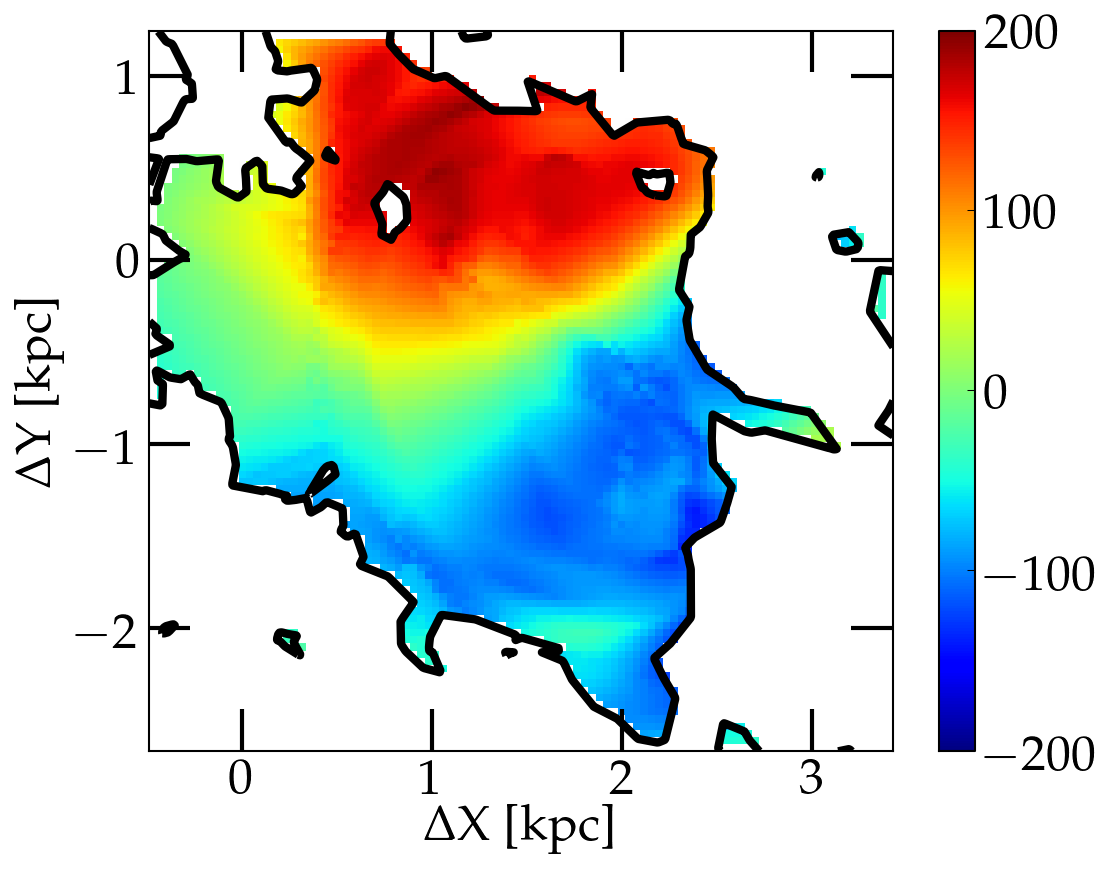}}
{\includegraphics[height=3.33cm,clip=true]
{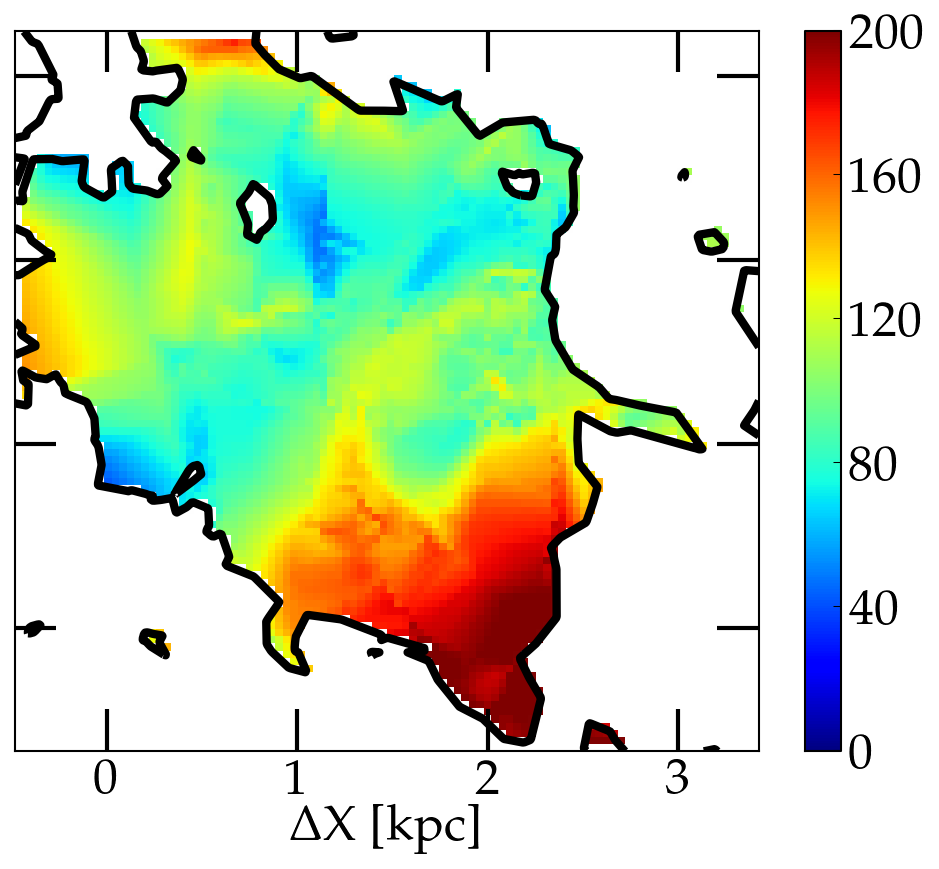}}
\caption{The un-smoothed first-moment (upper-left), second-moment (upper-right), first-moment uncertainty (middle-left) and second-moment uncertainty (middle-right) maps for the CO (5-4) transition. The lower panel shows the modelled (left) first- and (right) second-moment maps from our kinemetry analysis. The colour-bar is in units of km\,s$^{-1}$, and for the first-moment maps is shifted by $-$80~km\,s$^{-1}$ from a systemic velocity corresponding to a redshift of 3.042 (see Fig. \ref{fig:profiles}) using the radio definition of the velocity.}
\label{fig:CO5-4}
\end{center}
\end{figure}

\begin{figure}
\begin{center}
{\includegraphics[height=3.0cm, clip=true]
{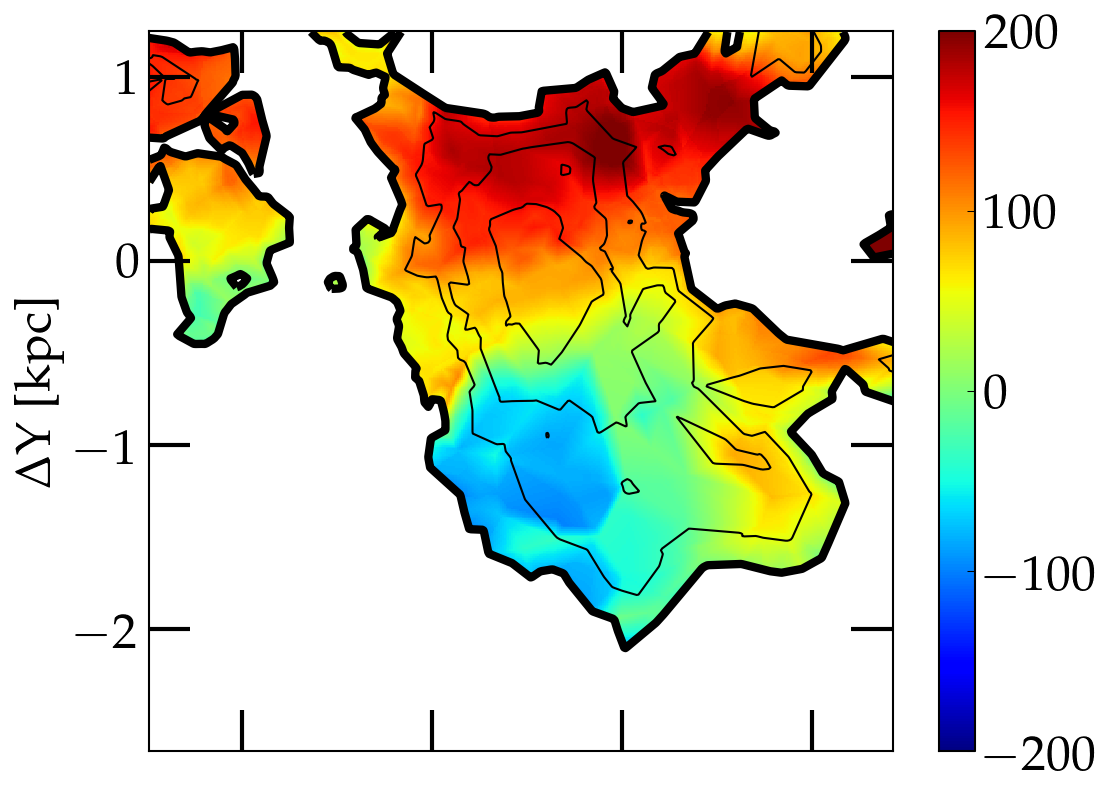}}
{\includegraphics[height=3.0cm, clip=true]
{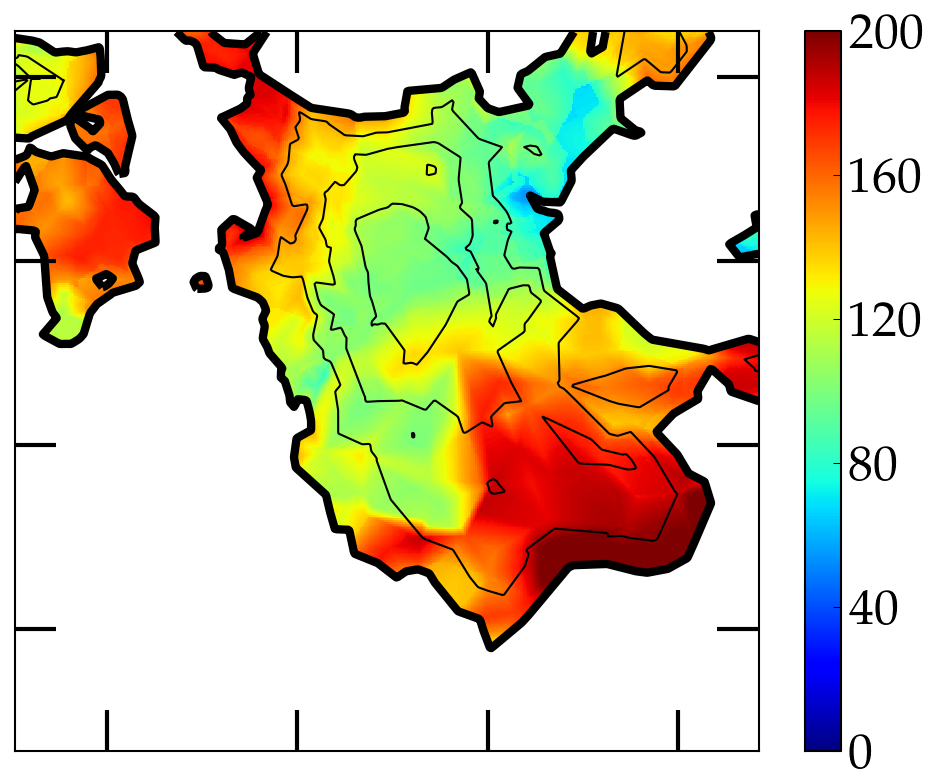}}\\
{\includegraphics[height=3.0cm, clip=true]
{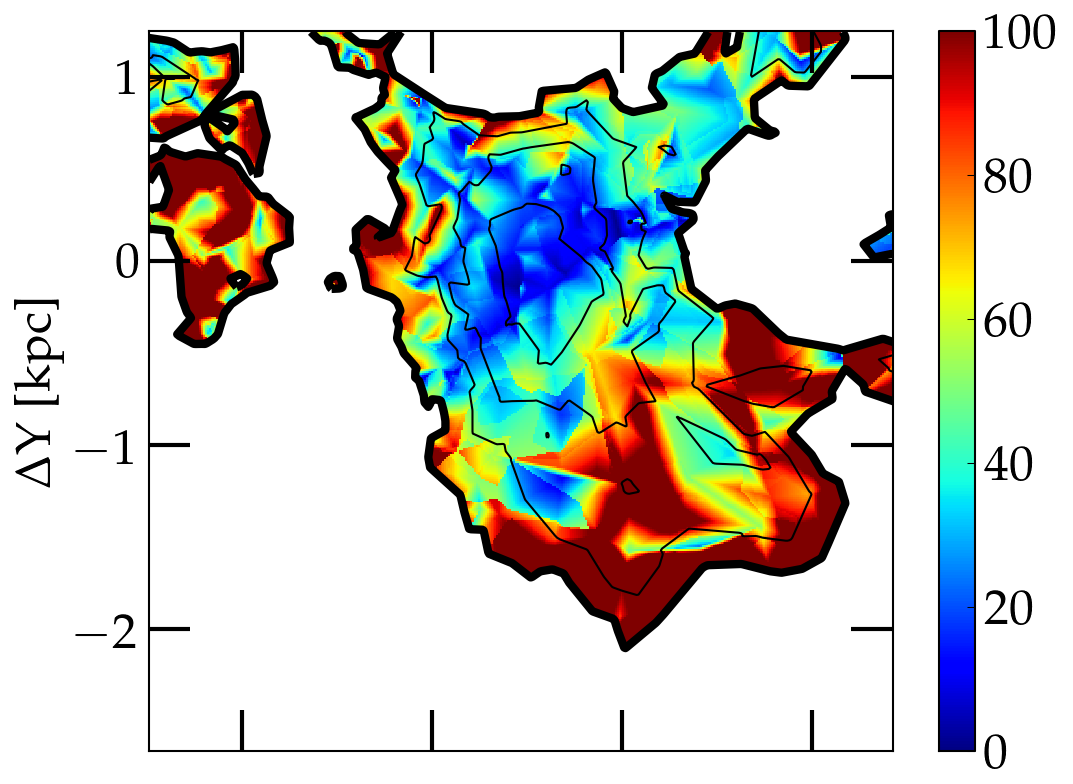}}
{\includegraphics[height=3.0cm, clip=true]
{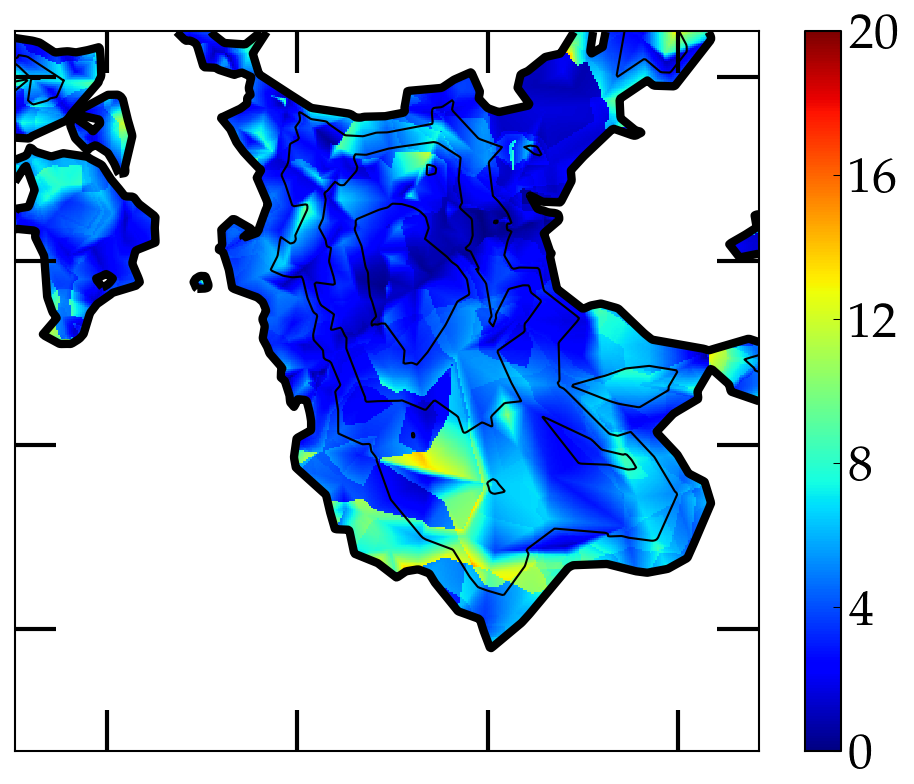}}\\
{\includegraphics[height=3.33cm, clip=true]
{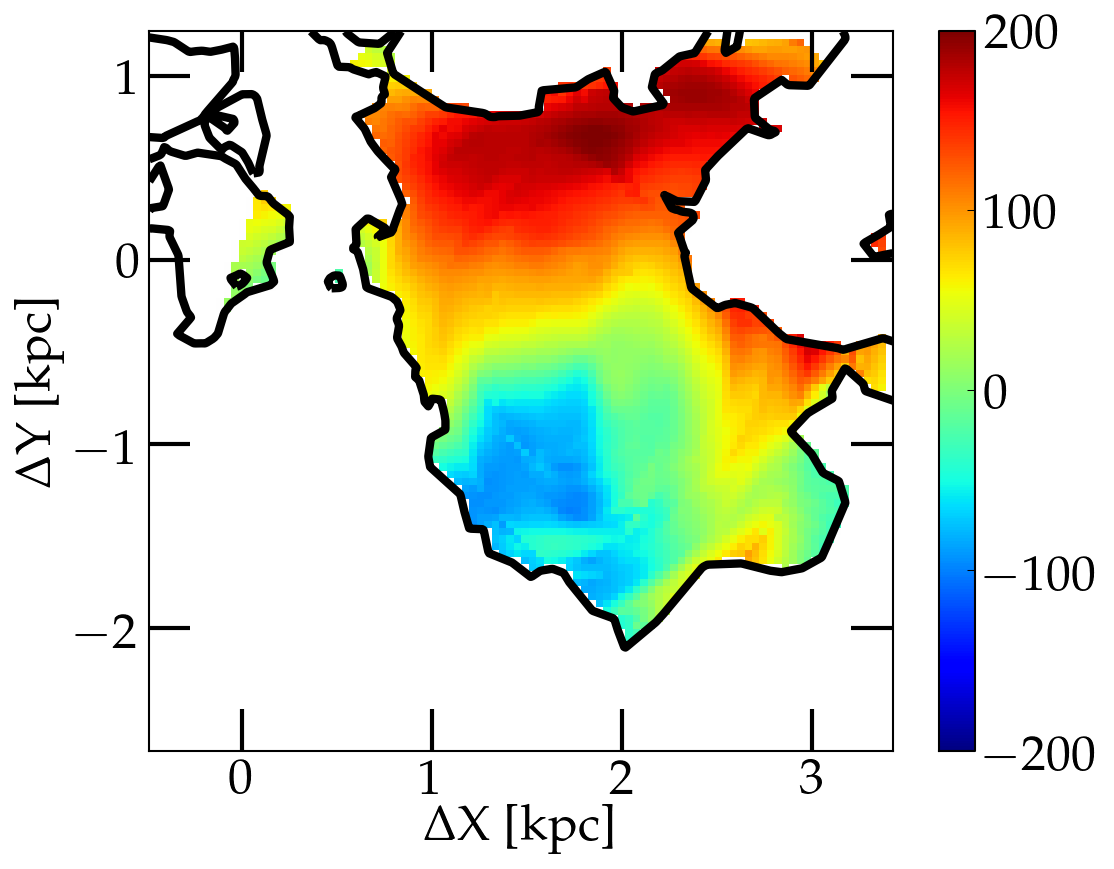}}
{\includegraphics[height=3.33cm, clip=true]
{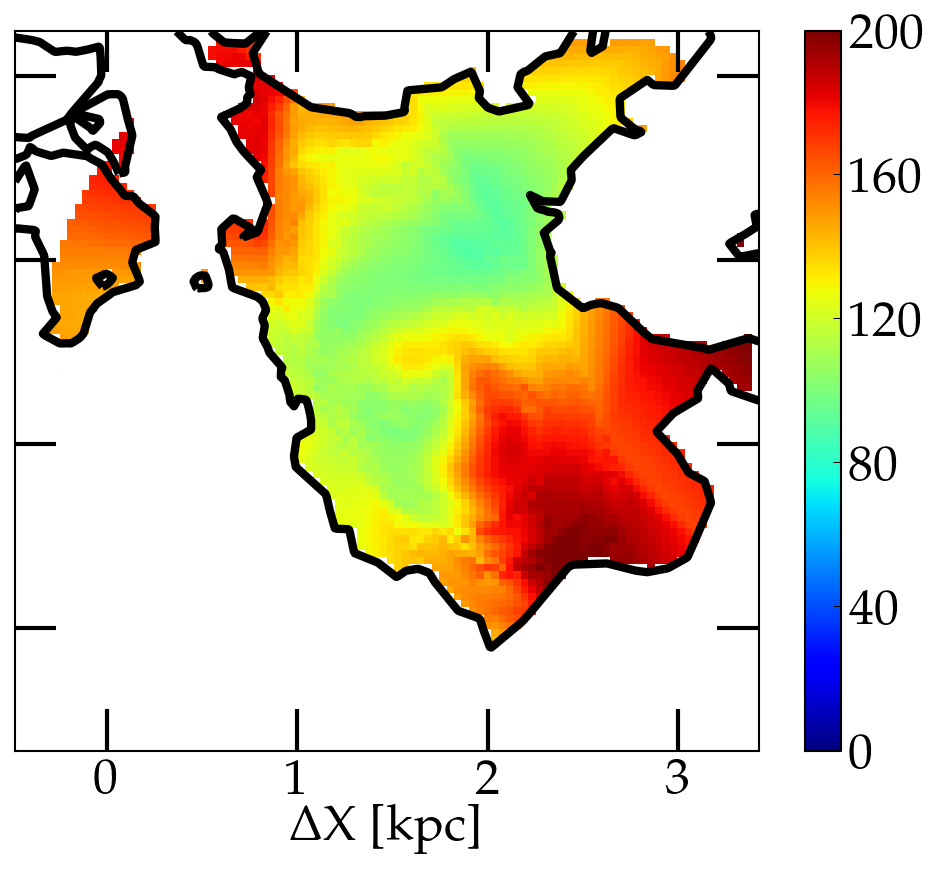}}
\caption{The same as for Fig.~\ref{fig:CO5-4}, but for the CO (8-7) transition.}
\label{fig:CO8-7}
\end{center}
\end{figure}

The first- and second-moment maps (velocity and velocity dispersion) of the CO (5-4) and CO (8-7) transitions are presented in Figs.~\ref{fig:CO5-4} and \ref{fig:CO8-7}, respectively. The velocity maps for both emission lines show a common central solid-body-like component in the north-south axis. However, the diffuse and extended component of the CO (5-4) and the extended tail of the CO (8-7) show a different velocity than expected for solid body rotation. 
Although the CO (5-4) extended feature has a slightly higher velocity dispersion than the rest of the emission, the velocity dispersion of the CO (8-7) tail feature is significantly different; the large $\sim$200~km\,s$^{-1}$ velocity dispersion is not due to a single broad velocity component, but is due to two velocity components along the same line-of-sight. We find that, the velocity dispersion of the CO (8-7) is overall higher than the CO (5-4) by about 40 km\,s$^{-1}$.

In order to quantify the asymmetry of the reconstructed velocity (dispersion) structure, we model the moment maps using the kinemetry method by \citet{{kranovic06}}. This is based on the idea that higher-order moments can be described by a series of concentric ellipses of increasing major-axis length; the moment profile along each ellipse is then decomposed into a Fourier series with a finite number of harmonic terms. From this analysis we find that both transitions are characterized by a significant level of rotation as well as a significant level of asymmetry at large radii. In the central regions of both transitions, the first harmonic term, $k _1$, which describes the rotational motion, is found to be substantial and the ratio $k_5/k_1$, which represents higher-order deviations from simple rotation, is consistent with zero, as expected for a disk-like rotation. However, in the outer regions, the $k_1$ term abruptly drops and at the same time the $k_5/k_1$ term significantly increases, as expected for multiple kinematical components or for post-coalescence perturbed disks \citep[][B12]{kranovic06,bellocchi12}. 

Using kinemetry, \citet[][S08]{shapiro08} have devised a  criterion to assess whether a galaxy is disk or merger dominated, based on the total kinematic asymmetry ($K_{\rm asym}$) in the dynamical state of the molecular gas. This criterion has been subsequently revised by B12, in order to better distinguish between isolated disks and post-coalescence mergers on the basis of large asymmetries at large radii. We quantify the total kinematic asymmetry to be $K_{\rm asym,S08} =$~0.65\,$\pm$\,0.10 and $K_{\rm asym,B12} =$~0.63$\pm$\,0.14 for the CO (5-4) transition and $K_{\rm asym,S08} =$~0.65\,$\pm$\,0.14 and  $K_{\rm asym,B12} =$~0.72\,$\pm$\,0.20 for the CO (8-7) transition (the uncertainties are a combination of the errors on the Fourier coefficients and the uncertainty due to the choice of centre).  These values are consistent with a merger/post-coalescence merger interpretation according to both criteria ($K_{\rm asym,S08}>$~0.50 and $K_{\rm asym,B12}>$~0.15). Note that although the classification criteria have been derived for data with angular resolutions that are much lower than the data presented here, we find that downgrading our resolution does not affect the derived $K_{\rm asym}$ values. Finally, the visual criteria by \citet{bellocchi13} would classify SDP.81 as a perturbed disk; while the velocity maps show a clear disk-like velocity gradient, the velocity dispersion map is relatively smooth without an obvious central peak, indicating that the disk is offset from the actual centre of mass. 

We estimate the dynamical mass of the disk structure by considering the maximum radius (1.56\,$\pm$\,0.07~kpc), maximum velocity (150\,$\pm$\,20~km\,s$^{-1}$) and the inclination angle (45\,$\pm$\,8~deg) of the outermost kinemetry ellipse that is consistent with an ordered rotation. We find an enclosed dynamical mass of $M$($<$\,1.56~kpc) = 1.6\,$\pm$\,0.6\,$\times$\,10$^{10}$~M$_{\odot}$, which although a factor $\sim$2 smaller than that determined by \citet{swinbank15}, is consistent at the 2$\sigma$-level. However, as the system is far from being in equilibrium, such dynamical mass estimates should be interpreted with care.

We find significant differences in our visibility-plane reconstructions (this letter; \citealt{Rybak15a}) and the image-plane reconstructions of \citet[][see also \citealt{swinbank15}]{dye15}. Our reconstructions detect more extended structure within the molecular gas distribution that are not seen in the image-plane analysis. Also, the position and intensity of the compact components are seen to vary between the two methods (note the compact structure varies significantly even within the individual image-plane analyses of the 1, 1.3 and 2~mm continuum data). These differences between the visibility- and image-plane results are likely due to the choice of weighting and deconvolution errors in the low SNR images of the image-plane channel data; the {\it uv}-taper and weighting that were applied to the image-plane data produces a point spread function with significant side-lobe structure that can introduce spurious artefacts in the channel images. For these reasons we believe that visibility-plane reconstructions are more robust and less biased, and the interpretation of intensity and velocity maps from image-plane reconstructions requires caution.

\section{Discussion \& Conclusions}
\label{conc}

\begin{figure}
\begin{center}
{\includegraphics[height=4.8cm, clip=true]
{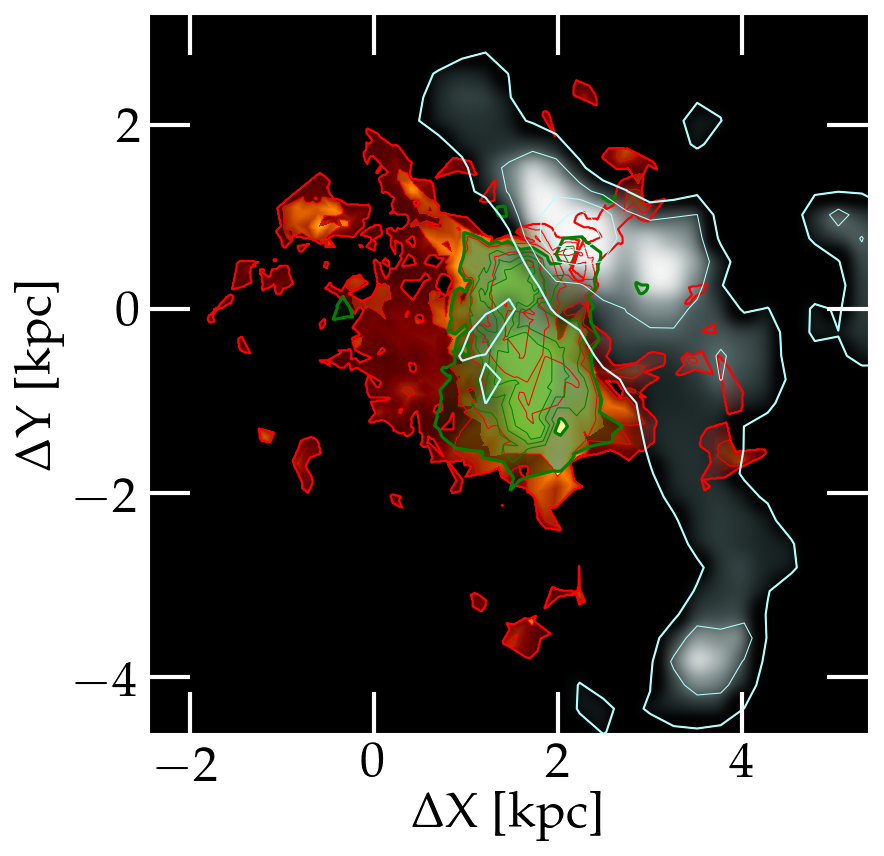}}
{\includegraphics[height=4.8cm, clip=true]
{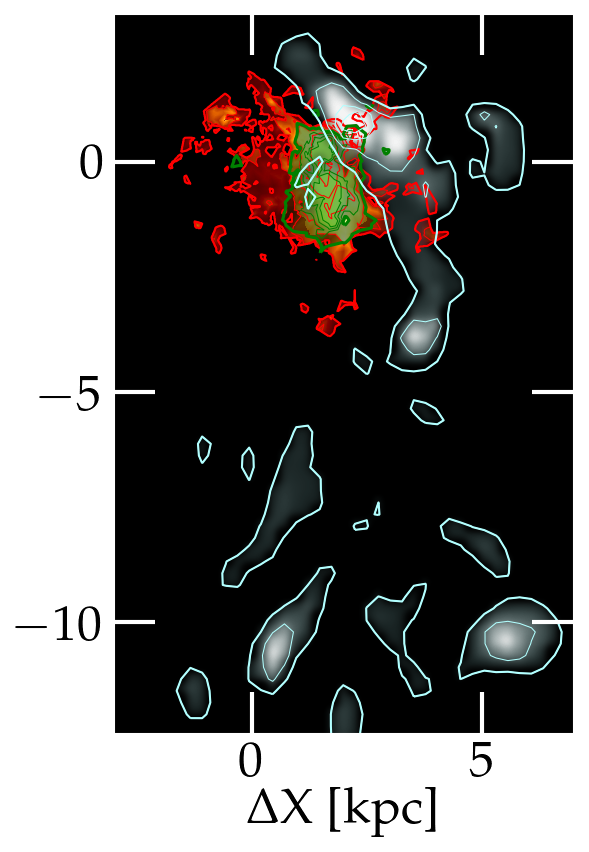}}
\caption{A composite source-plane reconstruction of the rest-frame FIR continuum dust (green; see Paper I), the combined CO (5-4) and CO (8-7) molecular gas (red) and the rest-frame UV/optical emission (grey).}
\label{fig:composite}
\end{center}
\end{figure}

The morphological state of SDP.81 is shown in Fig.~\ref{fig:composite}, where it is apparent that the reconstructed rest-frame UV/optical stellar component ({\it HST} F160W) and the combined CO (5-4) and CO (8-7) molecular gas component (both from this paper), and the heated dust from the rest-frame FIR continuum emission presented in Paper I form a complex structure. Taking the UV/optical stellar emission first, we see that this component is dominated by an elongated $\sim$4~kpc structure that is oriented by 45 degrees with respect to the north-south axis. There is another brightness peak about 4 kpc south of the main UV/optical peak, and they are connected by a low-surface brightness bridge. Note that the diffuse UV/optical structure that is also coincident with the CO (1-0) emission extends about $\sim$10~kpc south, as shown in Fig.~\ref{fig:composite}. The CO (5-4) and (8-7) molecular gas distribution also looks disturbed, with the region to the east being much more diffuse, whereas there is a sharp cut-off (shown by the small distance between the intensity contours) in the gas towards the west, which forms a ridge at the contact point with the stellar component. Interestingly, the extended tail that is seen in the CO (8-7) molecular gas distribution neatly fits into the area between the main UV/optical component and the bridge feature.

The ongoing star-formation, as traced by the heated dust emission, is concentrated to a region of about 2~kpc in size, and is coincident with the disk-like rotating structure within the inner part of the molecular gas distribution. This component also shows extension to the east and a sharp cut-off to the western edge. We note the detection of a compact molecular gas and star-formation component (20--60~M$_{\odot}$~yr$^{-1}$~kpc$^{-2}$) near the brightness peak of the UV/optical component, suggesting that star-formation is occurring in both structures. However, we find no further evidence of either heated dust or excited molecular gas emission across the full 15 kpc of the optical component, even though there is an established stellar population and a large reservoir of CO (1-0) available for star-formation. This suggests that the current burst of star-formation is localised to the northern part of the structure.

The central structure of SDP.81 is somewhat similar to GN20 (z = 4.05; \citealt{hodge12,hodge15}), where the UV/optical emission is offset from the molecular gas component, as traced by lower excitation CO (2-1) and the heated dust emission; these offsets are not uncommon in high-redshift galaxies and may be attributed to dust extinction of the UV/optical stellar emission or to strong negative stellar feedback together with gas accretion from gas clumps or satellites \citep{maiolino15}. Like GN20, there is also evidence for a central star-forming disk at the centre of SDP.81. However, there are also marked differences, which suggest that the starburst in SDP.81 may be within an interacting/merging system. First, our kinemetry analysis of the extended molecular gas distribution suggests a post-coalescence perturbed disk. Also, the UV/optical emission shows multiple compact and diffuse components over an elongated $\sim$15~kpc region. The CO (1-0) data in hand are not sensitive enough or have high enough angular resolution to determine the relative velocities of these individual components, but it is clear from the overall CO (1-0) emission that they form part of the same extended structure.

The merger scenario has been put forward for the triggering mechanism behind the strong star-formation observed within sub-mm galaxies. This is based on the observed multiple-component morphology and disturbed kinematics of unlensed starbursts and from simulations \citep[e.g.][]{Engel2010, narayanan10}. Indeed, this conclusion was also recently drawn by \citet{messias14} who studied the source plane properties of H-ALTAS~J1429$-$0028. For SDP.81, the situation is still not clear, but further observations of the CO (1-0) emission with the VLA at higher angular resolution could measure the relative dynamics of the various UV/optical components. This would confirm that they form a merging system, with large velocity offsets, or if they are part of a clumpy structure with a more chaotic velocity distribution. 

\section*{Acknowledgments}
This paper makes use of the following ALMA data: ADS/JAO.ALMA \#2011.0.00016.SV. ALMA is a partnership of ESO (representing its member states), NSF (USA) and NINS (Japan), together with NRC (Canada), NSC and ASIAA (Taiwan), and KASI (Republic of Korea), in cooperation with the Republic of Chile. The Joint ALMA Observatory is operated by ESO, AUI/NRAO and NAOJ. The National Radio Astronomy Observatory is a facility of the National Science Foundation operated under cooperative agreement by Associated Universities, Inc. Based on observations made with the NASA/ESA Hubble Space Telescope, obtained from the data archive at the Space Telescope Science Institute. STScI is operated by the Association of Universities for Research in Astronomy, Inc. under NASA contract NAS 5-26555.

\bsp

\label{lastpage}

\end{document}